\documentclass[a4paper,12pt]{article}
\usepackage[cp1251]{inputenc}
\usepackage[english,russian]{babel}
\usepackage[dvips]{graphicx}
\graphicspath{{.}}
\usepackage{amsfonts}
\usepackage{amsmath}

\begin{document}
\mathsurround=2pt \sloppy
\title{\bf Temperature dependence of the order parameter of the polar phase of liquid $^3$He in nematic aerogel}
\author{ I. A. Fomin
\vspace{.5cm}\\
{\it P. L. Kapitza Institute for Physical Problems}\\ {\it Russian
Academy of Science},\\{\it Kosygina 2,
 119334 Moscow, Russia}}
\maketitle
\begin{abstract}
It is shown that in the polar phase of superfluid $^3$He, stabilized by nematic aerogel, at the condition of specular reflection of quasi-particles from the strands of aerogel the temperature dependence of the gap in the spectrum of quasi-particles coincides with that in the bulk polar phase without impurities. The analogy with the Anderson theorem for conventional superconductors is discussed.

\end{abstract}

\section{Introduction}

 Experimental observation of the polar phase of superfluid $^3$He \cite{Dm-pp} and its further experimental investigation \cite{Dm-ms} became an important event in the physics of quantum fluids.All superfluid phases of $^3$He are formed as a result of Cooper pairing in the states with the orbital angular moment $l=$1 and spin $s=$1. The spin structure of the order parameter of the polar phase coinsides with that of the ABM-phase.  Specifics of the new phase manifests itself in its orbital part.he  polar phase corresponds to the state with projection of orbital moment $l_z=$0 on the direction of global anisotropy. The order parameter of the
 polar phase corresponds to one of minima of the free energy of superfluid $^3$He  \cite{VW}, but this phase is less favorable energetically than A and B- phases and it is unstable in the bulk liquid. V.V. Dmitriev and his co-workers \cite{Dm-pp}  succeeded in stabilization of the polar phase in a large volume with the aid of nematic  aerogel \emph{nafen} immersed in $^3$He. Nafen is a rigid structure \cite{nf} consisting of straight strands of  $Al_2O_3$  with the average diameter 8 - 9 nm  oriented nearly parallel to one direction.  The average distance between the strands for the samples of nafen used in the  experiments  \cite{Dm-pp}  varied from 18 to 64 nm. In these experiments the strands were covered by a film of $^4$He of 2.5$\div$3 atomic layers thick. This film prevented formation on the strands of a layer of solid paramagnetic $^3$He.  Possibility to stabilize the polar phase with the aid of nafen was confirmed by the experiments made in the other laboratories \cite{Parp,eltsov}. Even thou the method of stabilization of the polar phase by nafen has firm experimental justification its theoretical understanding is not that clear. The earlier attempts of Dmitriev`s group \cite{Dm-ob} to stabilyze the polar phase with the aid of another nematic aerogel (\emph{obninskii}) were not successful even thou the anisotropy of this aerogel was much higher than the anisotropy which according to the theoretical estimation \cite{AI} would be sufficient for this stabilization.    Further experiments  \cite{Dm-ms}  have shown, that  for  stabilization of the polar phase a character of scattering of quasiparticles on the strands of nafen is essential. In the experiments \cite{Dm-ms} the character of scattering could be changed by changing the thickness of the film of $^4$He, covering the strands of nafen . The polar phase was observed only when sufficiently thick $^4$He film was present. It is believed, that it makes scattering of quasiparticles close to specular at least at not too high pressures \cite{kj}.

Presently there is no universal microscopic theory, which would adequately descibe effects of different aerogels on the properties of superfluid 3He. The best is extention of the theory of superconducting alloys \cite{AG1,AG2} to the case of $p-wave$ Cooper pairing. which in some papers is dubbed as Homogeneous Scattering Model (HSM) \cite{thuneb}. This theory has a mean field character. It ignores fluctuations in a distribution of impurities and possible correlations between them. Neither of aerogels used in experiments with liquid 3He, and nafen in particular, meets conditions of applicability of the mean field theory. Predictions, based on the HSM model for that reason have only qualitative character. As a way to improve the situation less universal models were suggested, which take into account specific properties of the concrete aerogels \cite{thuneb}.

Specific property of nafen is its large anisotropy.  This property is taken into account by an idealized model \cite{i_f}, assuming that the strands are straight, mutually parallel and that they specularly reflect single particle excitations. Within this model longtudinal correlations in positions of impurities are taken into account exactly, they enter as conservation law for the longitudinal component of momenta of quasiparticles. An important result of this model is that transitions from the normal  phase of $^3$He filling space between the strands of nafen into the superfluid phases corresponding to the Cooper pairing with $l_z=0$ and $l_z=\pm$1 have to take place at different temperatures. The highest transition temperature corresponds to $l_z=0$, i.e. to the polar phase. Moreover, this temperature coinsides with the superfluid transition temperature of $^3$He without aerogel. This statement is analogous to one of the consequences of the theory of superconducting alloys \cite{AG2} for conventional Cooper pairing  known as \emph{Anderson theorem} \cite{And}. Anderson theorem includes even more strong statement that thermodynamic properties of conventional superconductors do not depend on a presence of non-magnetic impurities. In particular, these impurities do not change the temperature dependence of the superconducting gap in the spectra of single-particle excitations. Analogous statement for the polar phase has not been explicitely proven in Ref.\cite{i_f}, althou  it was implicitely used at the identification of the phase \cite{Dm-pp}. It was used also at the experimental proof  of the existence of line of nodes in the gap for the polar phase \cite{eltsov}.  In the present paper a different from Ref.\cite{i_f}  argument is used, which includes a formal proof of independence of the temperature behavior of the amplitude of the superfluid gap in the polar phase of $^3$He on a presence of nafen for the previously formulated model. That resumes a full proof of the analog of Anderson teorem for the polar phase.

\section{Argument}
Theory of Abrikosov and Gorkov (AG in what follows) \cite{AG1,AG2} was applied to the p-wave triplet Cooper pairing by Larkin \cite{Lark}. We follow his argument. As the order parameter in this theory is used the anomalous average $F_{\alpha\beta}(\hat{k})$, which at a triplet pairing is 2$\times$2 symmetric spin matrix depending on a direction  $\hat{k}$ in the momentum space. It can be written as a combination of Pauli matrices $\sigma^{\mu}_{\alpha\beta}$:  $F_{\alpha\beta}(\hat{k})\sim d_{\mu}(\hat{k})\sigma ^{\mu}_{\alpha\beta}\sigma^y$. Index $\mu$ runs over 3 values: $x,y,z$. Polar phase is one of the Equal Spin Pairing (ESP) phases for which direction of $d_{\mu}$  does not depend on $\hat{k}$. Effect of orientation of $d_{\mu}$ on the amplitude of the gap stems from a small dipole-dipole interaction and can be neglected. It is convenient to orient  $d_{\mu}$ along $y$ axis, then $F_{\alpha\beta}(\hat{k})\sim\delta_{\alpha\beta}$ and basic equations of the theory can be written as the scalar equations for the normal $G(\hat{k})$ and anomalous $F^{\dag}(\hat{k})$ Green functions:
$$
(i\omega_n-\xi-\overline{G_{\omega}}(\hat{k}))G(\hat{k},\omega_n)+(\Delta(\hat{k})+\overline{F_{\omega}}(\hat{k}))F^{\dag}(\hat{k},\omega_n)=1,  \eqno(1)
$$
$$
(i\omega_n+\xi+\overline{G_{-\omega}}(\hat{k}))F^{\dag}(\hat{k},\omega_n)+(\Delta^*(\hat{k})+\overline{F^{\dag}_{\omega}}(\hat{k}))G(\hat{k},\omega_n)=0.  \eqno(2)
$$
Self energies $\overline{G_{\omega}}(\hat{k})$, $\overline{F_{\omega}}(\hat{k})$ and $\overline{F^{\dag}_{\omega}}(\hat{k})$ depend on a potential of interaction of quasiparticles with the impurities. It is assumed in the idealized model of nafen \cite{i_f} that potential of this interaction does not depend on the coordinate $z$ along the direction of the strands:  $U(\mathbf{r})=\sum_a u(\rho-\rho_a)$, where  $\rho=(x,y)$ is two-dimensional vector. Index $a$ enumerates the strands. The equations will contain Fourier transform of the potential:
$$
U(\mathbf{k})=2\pi\delta(k_z)u(\kappa)\sum_a e^{-i\kappa\rho_a},                              \eqno(3)
$$

where  $\kappa=(k_x,k_y)$ is a two-dimensional wave vector and $u(\kappa)=\int u(\rho)\exp(i\kappa\rho)d^2\rho$.
For this model
$$
\overline{G_{\omega}}(\hat{k})=n_2\int|u(\kappa-\kappa_1)|^2G(\kappa_1,k_z)\frac{d^2\kappa_1}{(2\pi)^2},                       \eqno(4)
$$
$$
\overline{F^{\dag}_{\omega}}(\hat{k})=n_2\int|u(\kappa-\kappa_1)|^2F^{\dag}(\kappa_1,k_z)\frac{d^2\kappa_1}{(2\pi)^2}.                  \eqno(5)
$$
Here $n_2$ is the two-dimensioal density of strands.

The system (1)-(6) differs from that, considered in Ref. \cite{Lark} by the form of the potential of impurities (4). Presence of $\delta(k_z)$ in this potential means that $k_z$ is conserved at the scattering of quasiparticles by the strands. Because of this conservation Eqns. (1)-(5) have solution $\Delta(\hat{k})=\Delta(T)\hat{k}_z$, which does not contain  contribution of other projections of $\hat{k}$. It has been shown \cite{i_f} that this solution describes the phase with the highest transition temperature from the normal phase.
To find the explicit form of this solution one has to split $\overline{G_{\omega}}(\hat{k})$ into even and odd parts over $\omega$: $g_e(\omega,\hat{k})=\frac{1}{2}[\overline{G_{\omega}}(\hat{k})+\overline{G_{-\omega}}(\hat{k})]$ and  $g_o(\omega,\hat{k})=\frac{1}{2}[\overline{G_{\omega}}(\hat{k})-\overline{G_{-\omega}}(\hat{k})]$, and after that to introduce new variables:  $i\widetilde{\omega}_n=i\omega_n-g_o(\omega,\hat{k})$, $\tilde{\xi}=\xi+g_e(\omega,\hat{k})$, $\widetilde{\Delta}^{\dag}=\Delta^{\ast}+\overline{F^{\dag}_{\omega}}(\hat{k})$.
 In terms of these variables Eqns. (1), (2) have a simple form:
$$
(i\widetilde{\omega}_n-\tilde{\xi})G(\hat{k},\omega_n)+\tilde{\Delta}F^{\dag}(\hat{k},\omega_n)=1,                                      \eqno(6)
$$
$$
\widetilde{\Delta}^{\dag}G(\hat{k},\omega_n)+(i\widetilde{\omega}_n+\tilde{\xi})F^{\dag}(\hat{k},\omega_n)=0.                        \eqno(7)
$$
These equations are easily solved vith respect to $G(\hat{k},\omega_n)$ and $F^{\dag}(\hat{k},\omega_n)$:
$$
G(\hat{k},\omega_n)=-\frac{i\widetilde{\omega}_n+\tilde{\xi}}{(\widetilde{\omega}_n)^2+\tilde{\xi}^2+|\widetilde{\Delta}|^2},\qquad F^{\dag}(\hat{k},\omega_n)=\frac{\widetilde{\Delta}^{\dag}}{(\widetilde{\omega}_n)^2+\tilde{\xi}^2+|\widetilde{\Delta}|^2}.                     \eqno(8)
$$
Substitution of these expressions in the Eqns. (4),(5) renders equations for $\overline{G}$ and $\overline{F^{\dag}}$ and their combinations:
$$
g_e(\omega,\hat{k})=-n_2\int|u(\kappa-\kappa_1)|^2\frac{\tilde{\xi}}{(\widetilde{\omega}_n)^2+\tilde{\xi}^2+|\widetilde{\Delta}|^2}\frac{d^2\kappa_1}{(2\pi)^2}.  \eqno(9)
$$
Integral in the r.h.s. diverges at large  $\tilde{\xi}$, i.e.a principal contribution to the integral comes from the states, which are far from the Fermi surface. Like in the case of $s$-wave pairing the increment $g_e(\omega,\hat{k})$ can be absorbed in renormalization of the chemical potential. Then $\tilde{\xi}$ is a new variable of integration, counted from the renormalized Fermi energy.
Temperature dependence of the gap is determined by the equation:
$$
\Delta^{\dag}(\hat{k})=-T\sum_n\sum_{\mathbf{k}}V(\mathbf{k},\mathbf{k'})F^{\dag}(\hat{k},\omega_n).             \eqno(10)
$$
 The pairing interaction in Eq. (10) is usually taken in a form
$$
V(\mathbf{k},\mathbf{k'})=3g(\hat{k}\cdot\hat{k'}).              \eqno(11)
$$
Eq. (10) is a closed equation for the gap if the function $F^{\dag}(\hat{k},\omega_n)$ is expressed in terms of original variables $\omega_n$  and  $\Delta^{\dag}(\hat{k})$.
For the idealized model of nafen it is possible to make a substitution  $i\widetilde{\omega}_n=i\omega_n\eta(k_z,\omega_n)$, $\widetilde{\Delta}^{\dag}=\Delta^{\dag}\eta(k_z,\omega_n)$ with one function $\eta(k_z,\omega_n)$:
$$
\eta(k_z,\omega_n)=1+\frac{m^*n_2}{4\pi}\int\frac{|u(\kappa-\kappa_1)|^2d\varphi_1}{\sqrt{\omega_n^2+|\Delta|^2}}.               \eqno(12)
$$
Integration here goes over the angle $\varphi_1$ between $\kappa_1$ and $\kappa$ and $m^*$is the effective mass of quasiparticles. Substitution of $\overline{F^{\dag}}$ in Eq. (10) renders:
$$
\Delta^{\dag}(\hat{k})=-T\sum_n\sum_{\mathbf{k'}}3g(\mathbf{k}\cdot\mathbf{k'})\frac{\Delta^{\dag}(\hat{k'})\eta}{\omega_n^2\eta^2+|\Delta(\hat{k'})|^2\eta^2+\xi'^2}.             \eqno(13)
$$
Summation over $\mathbf{k'}$ here has to be changed for the integration and  the momentum dependent solution
$\Delta(\hat{k})=\Delta(T)\hat{k}_z$ has to be substituted.
After rescaling $\xi'=vT\eta$ the function $\eta$ remains in the equation for $\Delta(T)$ only in the renormalization of the cut-off energy. Such renormalization introduces corrections of the order of $1/lk_F$, which are beyond the accuracy of the present argument. Integration over $\varphi'_1$ renders  zero contribution for the terms of the scalar product $(\hat{k}\cdot\hat{k'})$ which contain the transverse components $\hat{k'}_x$,   $\hat{k'}_y$ and the overall factor 2$\pi$ for other terms.
Neglecting small corrections we arrive at the equation which does not contain the cross-section of scattering of quasi-particles by the strands of nafen:
$$
1=\frac{3gm^*k_F}{\pi^2}\int_0^1 x^2dx\sum_n\int_0^{v_{max}}dv\frac{1}{(2n+1)^2+v^2+(\bar{\Delta}(T)x)^2},   \eqno(14)
$$
where $\bar{\Delta}=\Delta(T)/(\pi T)$. Eq. (14) coinsides with the equation, detrmining temperature dependence of the gap in the bulk polar phase of superfluid  $^3$He. The cut-off $v_{max}$ can be expressed through the observable parameters $T_c$ or $\Delta_0=\Delta(T=0)$. There are standard procedures of analysis of Eq. (14) in the limits $T\rightarrow T_c$ and $T\rightarrow 0$ \cite{AGD}.

At $T\rightarrow 0$ the ratio in the r.h.s. of Eq. (14) has to be expanded in powers of $(\bar{\Delta}(T)\hat{k}_z)^2$. Following practically literally the argument of the Ref. \cite{AGD} we arrive at the asymptotic expansion
$$
\ln\frac{T}{T_c}=-\frac{3}{5}\frac{7}{8}\zeta(3)(\frac{\Delta}{\pi T})^2+\frac{3}{7}\frac{93}{128}\zeta(5)(\frac{\Delta}{\pi T})^4-... ,         \eqno(15)
$$
where $\zeta(z)$ is Riemann $\zeta$-function. Eq. (15) differs fron the corresponding equation for the s-wave case only by extra coefficients 3/5 and 3/7 in front of  $(\bar{\Delta}(T)\hat{k}_z)^2$ and $(\bar{\Delta}(T)\hat{k}_z)^4$ respectively. The difference is due to the angular dependence of the gap. The $T_c$
 here is the temperature of the superfluid transition in the bulk liquid $^3$He \cite{i_f}.

 In the limit  $T\rightarrow 0$ it is more convenient first to perform summation in Eq. (14) over frequencies and use as a starting point the equation
$$
1=\frac{3gm^*k_F}{2\pi^2}\int_0^1 x^2dx\int_0^{v_{max}}\frac{dv}{\sqrt{v^2+(\bar{\Delta}x)^2}}\tanh\frac{\sqrt{v^2+(\bar{\Delta}x)^2}}{2},   \eqno(16)
$$
 where $x=k_z$. Asymptotic solution of this equation at  $T\rightarrow 0$ is discussed in detail in the Supplementary material of Ref.\cite{eltsov}.

\section{Discussion}
The original Anderson theorem applies to a conventional (s-wave) superconductor with  the ensemble of non-magnetic impurities. The statement of the theorem is eventually based on the $t\rightarrow -t$ symmetry of the system, i.e. on the fact that the states of electrons, forming the Cooper pair are transformed into each other at the reversal of time. This requirement is met exactly for every realization of the ensemble. The discussed above analog of Anderson theorem for the polar phase applies to the idealized model of nafen. The crucial property of this model is the specular reflection of quasiparticles by the strands of nafen and ensuing conservation of the longitudinal projection of their momenta. This property also holds exactly for all realizations of the idealized system but may not hold for real nafen.

The unexpected reaction of some superconductors on addition of impurities was recently discussed in connection with  \emph{multi-orbital}  superconductors \cite{fisher,ramires1,art}. The superfluid $^3$He falls in this category. In case of $^3$He it means that Cooper pairs have orbital moment $l=1$, but they can be distinguished by additional quantum number - its projection $l_z=0,l_z=\pm 1$. In the bulk liquid transition temperature $T_c$ is the same for all three projections. In the natural basis $l_z=0,l_z=\pm 1$  potential of impurities (4) is block-diagonal. In terms of Ref. \cite{fisher} it means, that for the polar phase only \emph{intra-band} Cooper pairing is presnt and this phase meets the simple criterion of stability.
Admixture of non-specular scattering introduces possibility of \emph{inter-band} Cooper pairing.  It has a detrimental effect on the formation of Cooper pairs in addition to possible   magnetic scattering. A coupling between the state with $l_z=0$ and states with $l_z=\pm 1$ can bring in  admixture of the transverse  components to the order parameter of the superfluid phase with the highest transition temperature and to lower this temperature in comparison with the $T_c$ for the bulk liquid. It also breaks the axial symmetry of the superfluid phase which is formed just below the $T_c$.

This paper is written for the special issue of JETP, devoted to 100-th anniversary of academician A.S. Borovik-Romanov. In the 80-ies of the last sentury I had opportunity to collaborate with the experimental group of the P.L. Kapitza Institute, led by Andrey Stanislavovich. The group included also Yu.M. Bunkov, V,V. Dmitriev and Yu.M. Mukharskii. It was real collaboration. We worked with enthusiasm. Frequent discussions of the current results, which we all, including Andrey Stanislavovich held in the laboratory are still of my best recollections of this time.

I thank G.E. Volovik, who brought to my attention the papers \cite{fisher,ramires1,art} and V,V. Dmitriev for useful comments.

\end{document}